\documentclass{emulateapj}
\usepackage{natbib} 
\usepackage{graphicx,verbatim}
\usepackage{bm}
\usepackage{hyperref}
\usepackage{color,graphics,amsmath,amssymb}

\def\fig{Figure~}

\def\lcdm{$\Lambda$CDM~}
\def\healpix{HEALpix~}
\def\cl{C_{\ell}}
\def\apj{ApJ}
\def\mnras{MNRAS}
\def\nside{N_{\mathrm{side}}}

\def\mcb{\mathcal{C}_b}
\def\mcbp{\mathcal{C}_{b'}}

\def\lmax{\ell_{\mathrm{max}}}
\def\cosmomc{CosmoMC~}

\newcommand{\beq}{\begin{equation}}
\newcommand{\eeq}{\end{equation}}

\begin{document}

\title{Directional dependence of \lcdm cosmological parameters}
\author{M. Axelsson\altaffilmark{1}, Y. Fantaye\altaffilmark{1}, F. K. Hansen\altaffilmark{1}, A. J. Banday\altaffilmark{2}, H. K. Eriksen\altaffilmark{1}, K. M. Gorski\altaffilmark{3,4,5}}

\altaffiltext{1}{Institute of Theoretical Astrophysics, University of Oslo,
P.O. Box 1029 Blindern, N-0315 Oslo, Norway} 

\altaffiltext{2}{Universite de Toulouse, UPS-OMP, IRAP, Toulouse, France}

\altaffiltext{3}{Jet Propulsion Laboratory, M/S 169/327, 4800 Oak Grove Drive,
Pasadena CA 91109}
\altaffiltext{4}{Warsaw University Observatory, Aleje Ujazdowskie
4, 00-478 Warszawa, Poland}
\altaffiltext{5}{California Institute of Technology,Pasadena CA 91125} 

\email{magnus.axelsson@astro.uio.no}
\email{y.t.fantaye@astro.uio.no}

\begin{abstract}
  We study hemispherical power asymmetry
  \citep{eriksen2004,hansen2004,hansen2009} in the WMAP 9-year
  data. We analyse the combined V- and W-band sky maps, after
  application of the KQ85 mask, and find that the asymmetry is
  statistically significant at the $3.4\sigma$ confidence level for
  $\ell=2$--600, where the data is signal dominated, with a preferred
  asymmetry direction $(l,b)=(227,-27)$. Individual asymmetry axes
  estimated from six independent multipole ranges are all
  consistent with this direction.  Subsequently, we estimate
  cosmological parameters on different parts of the sky and show that
  the parameters $A_s$, $n_s$ and $\Omega_b$ are the most sensitive to
  this power asymmetry. In particular, for
  the two opposite hemispheres aligned with the preferred asymmetry
  axis, we find $n_s = 0.959 \pm 0.022$ and $n_s = 0.989 \pm 0.024$,
  respectively.
\end{abstract}



\section{Introduction}

Shortly after the release of the first-year WMAP data
\citep{bennett:2003}, \citet{eriksen2004} and \citet{hansen2004}
reported a detection of a hemispherical power asymmetry in the cosmic
microwave background (CMB) on large angular scales in the multipole
range $\ell=2$--40.  The power in this multipole range was found to be
significantly higher in the direction towards Galactic longitude and
latitude $(l=237^\circ, b=-20^\circ)$ than in the opposite direction.
Due to computational limitations at the time, higher multipoles were
not investigated. These findings were supported by numerous other
studies, e.g., \citet{minkowski, hansen2009} and references therein.
However, the significance of the results has often been called into
question, in particular, due to the alleged a-posteriori nature of the
statistics used.  In particular, it is debated whether the statistic
has been designed to focus on visually anomalous features revealed by
an inspection of the data \citep[e.g.,][]{bennett:2011}.

The only rigorous way to contend with this assertion is by performing
repeated experiments and analysing the resulting independent data sets
that may provide additional information. For cosmological studies,
this is in general difficult, given that there is only one available
Universe. However, it is not impossible --- the standard inflationary
cosmological model
assumes that the Universe is homogeneous and isotropic, and that the
initial fluctuations have amplitudes that are Gaussian distributed,
independent and with random phase.  This implies that different
physical scales should be statistically uncorrelated, and therefore
the morphology of the largest scales should not have any predictive
power over the morphology of the smaller scales. For the power
asymmetry, this suggests that there is a possibility to study
effectively new data sets by considering angular scales that have not
previously been studied.

This extension to smaller angular scales was first undertaken by
\citet{hansen2009}, when analyzing the WMAP 5-year temperature data
set. The asymmetry was then found to extend over the range
$\ell=2-600$ with a preferred direction $(l=226^\circ, b=-17^\circ)$
for the higher multipoles, fully consistent with the direction for the
lower multipoles found in the original 1-year WMAP analysis. Two
approaches were used for the analysis: (1) a statistical model
selection procedure taking into account the penalty for including 3
new parameters (amplitude and direction of asymmetry), which showed
that indeed an asymmetric model was preferred;
and (2) a simple test in which the preferred power asymmetry axis was
estimated independently for six multipole bins of width
$\Delta\ell=100$.
It was found that these directions, which should be statistically
independent, were strongly aligned; none of the 10\,000 simulated
isotropic CMB maps showed a similarly strong clustering of preferred
directions.  An alternative approach modeled the power asymmetry in
terms of a dipolar modulation field, as suggested by
\citet{gordon:2005}. \citet{hoftuft2009} found a $3.3\sigma$ detection
using data smoothed to an angular resolution of $4.5^{\circ}$ FWHM,
with an axis in excellent agreement with previous results. These
studies, covering very different angular scales than those used in the
original analysis, argue against an a-posteriori interpretation of the
effect.

In this Letter, we repeat the high-$\ell$ analysis due to
\citet{hansen2009} using the WMAP\footnote{http://www.lambda.gsfc.nasa.gov} 
9-year data (hereafter referred to
as WMAP9, with a similar notation for the first- and five-year data sets).
However, the main goal is to estimate cosmological parameters in
the two maximally asymmetric hemispheres, in order to assess their
stability with respect to the power asymmetry \citep[for a closely related
theoretical study, see][and references therein.]{moss:2011}.
A similar analysis was performed in \citet{asymparam} using
the WMAP first-year data, but only taking into account the asymmetry
observed in the $\ell=2-40$ range, limited by a grid-based approach,
and consequently only considering a few cosmological parameters. In
the following, we use CosmoMC\footnote{COSMOlogical Monte Carlo software package 
(http://www.cosmologist.info/cosmomc).} 
to obtain the full posterior of all relevant cosmological parameters using the 
entire multipole range afforded by the WMAP9 data. { 
We adopt canonical \lcdm as our baseline cosmological model, with six
parameters - 
the baryon density today $\Omega_b h^2$, 
the Cold dark matter density today $\Omega_{DM} h^2$, 
the scalar spectrum power-law index $n_s$,
the log power of the primordial curvature perturbations $log[10^{10} A_s]$, 
the angular size of the sound horizon at recombination $\theta$, 
and the Hubble constant $H_0$, where $h$ represents this value in
units of 100\,km\,s$^{-1}$\,Mpc$^{-1}$.
}  


\section{Data and Method}
\label{sect:data}

\begin{figure}
\includegraphics[width=\hsize]{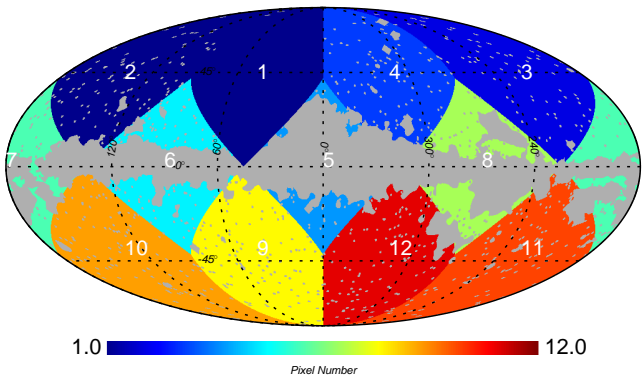}
\caption{The 12 sky patchs used in this paper: the regions are
  delineated by the intersection of the 12 HEALPix base pixels with
  the WMAP9 KQ85 mask.}
\label{fig:regions}
\end{figure} 

\begin{figure}
\centering
\includegraphics[width=\hsize]{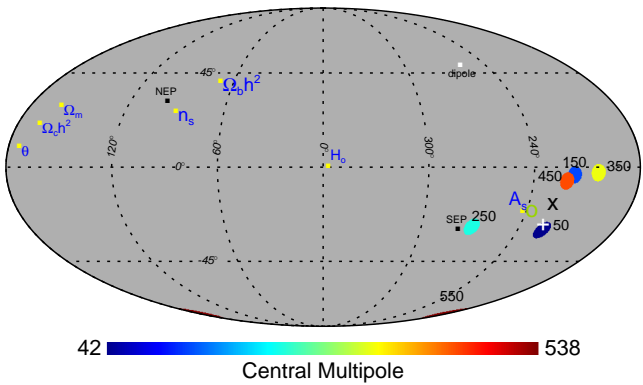}
\caption{Dipole directions for maps of the local power spectrum
  computed for the 12 regions in Figure~\ref{fig:regions} 
  from the WMAP9 combined V- and W-band data and 
  separated into six 100-multipole bins. We also show the direction for the full $\ell=2-600$
  range (white cross), for the $\ell=2-40$ interval determined from
  WMAP1 (green circle) and the $\ell=2-600$ range
  from WMAP5 (black cross). NEP and SEP denote the North and South
  Ecliptic Poles, respectively. The dipole directions for the local
  parameter estimate maps are also shown.
}
\label{fig:dipoles}
\end{figure}

\begin{figure}
\centering
\includegraphics[width=\hsize]{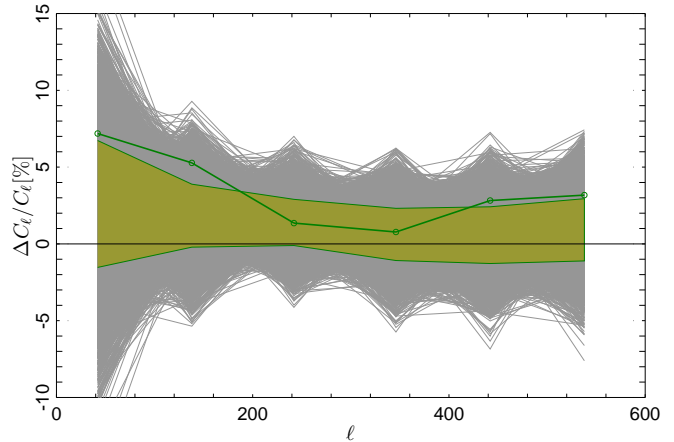}
\caption{The ratio of the local power spectra computed from
  antipodal hemispheres centred along the preferred
  dipole direction, as determined over the $\ell=2-600$ range. The
  connected dark-green open circles correspond to the values derived
  from the WMAP9 coadded V- plus W-band data. The grey
  lines are the individual power spectra ratios as computed
  for the preferred dipole direction of each of the 10\,000
  simulations. The olive-green band with dark-green bounding curves
  represents the corresponding 68$\%$ confidence levels.}
\label{fig:cldiff}
\end{figure} 

We use the publicly available WMAP9 temperature sky maps
\citep{wmap91}, co-adding (with inverse-noise-variance weighting) the
V (61~GHz) and W (94~GHz) band foreground-cleaned maps. We also
generate a set of 10\,000 simulated CMB-plus-noise Monte Carlo (MC) simulations based
on the WMAP best-fit $\Lambda$CDM power spectrum \citep{wmap92}, noise
rms maps and beam profiles for the V and W bands. The WMAP9 KQ85
Galactic and point source mask is used to remove pixels with high
foreground contamination.


\paragraph{Power asymmetry: } 
{ The MASTER \citep{hivon2002} approach is used to estimate the
  power spectra, $\cl$, from pseudo-spectral estimators applied to
  local regions of the sky. When computing the MASTER kernel, we bin
  the pseudo-spectra into bins of width $\Delta \ell = 16$ in order to
  avoid a singular matrix. This version of the spectra is used later
  for parameter estimation.}

{ In order to estimate the dipole directions of the local spectra,
  we first obtain an $\nside=1$ map, as illustrated in
  \fig~\ref{fig:regions}, where the value of each pixel is the binned
  power spectrum for a given range in $\ell$. However, in this case we
  combine the $16\ell$-bins further into blocks containing
  approximately 100 multipoles, following the procedure used in
  \citet{hansen2009}, and thereby reducing the uncertainty on the
  direction. From this map we then estimate the dipole amplitude and
  direction using an inverse variance weighting of the pixels; the
  variance of each pixel is calculated using 10\,000 isotropic
  simulations which incorporate the noise and beam properties 
  of the data, and to which the same mask has been applied. }






\begin{figure*}[ht!]
\centering
\includegraphics[width=0.8\textwidth,angle=0]{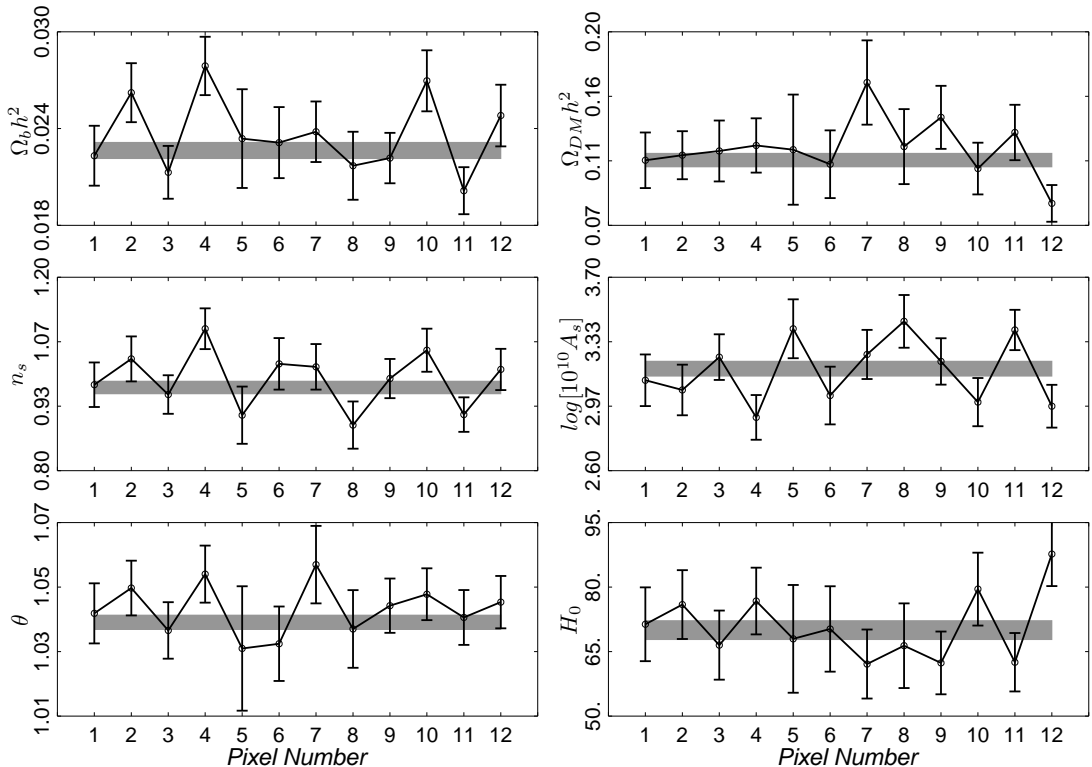}
\caption{Estimated \lcdm parameter values for the 12 regions defined
  in Figure~\ref{fig:regions}, as indicated by the open circles. The estimates are based on
  multipoles in the range $\ell = 2-1008$. The grey band represents the 68\%
  confidence level determined from a likelihood analysis of the WMAP9 data with the KQ85 mask applied.}
\label{fig:lcdmpat}
\end{figure*}

\begin{figure}
\includegraphics[width=\linewidth,angle=0]{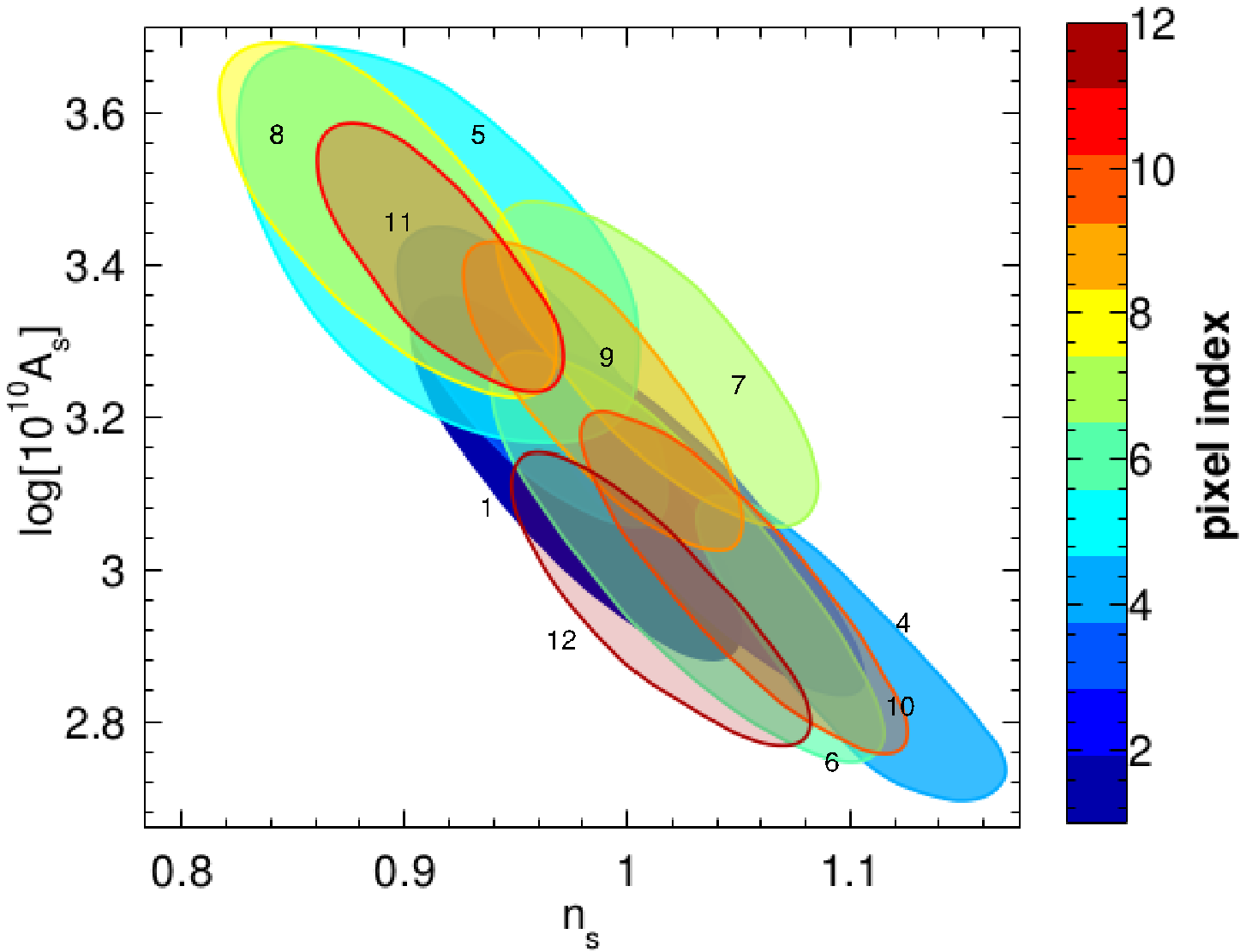}

\includegraphics[width=\linewidth,angle=0]{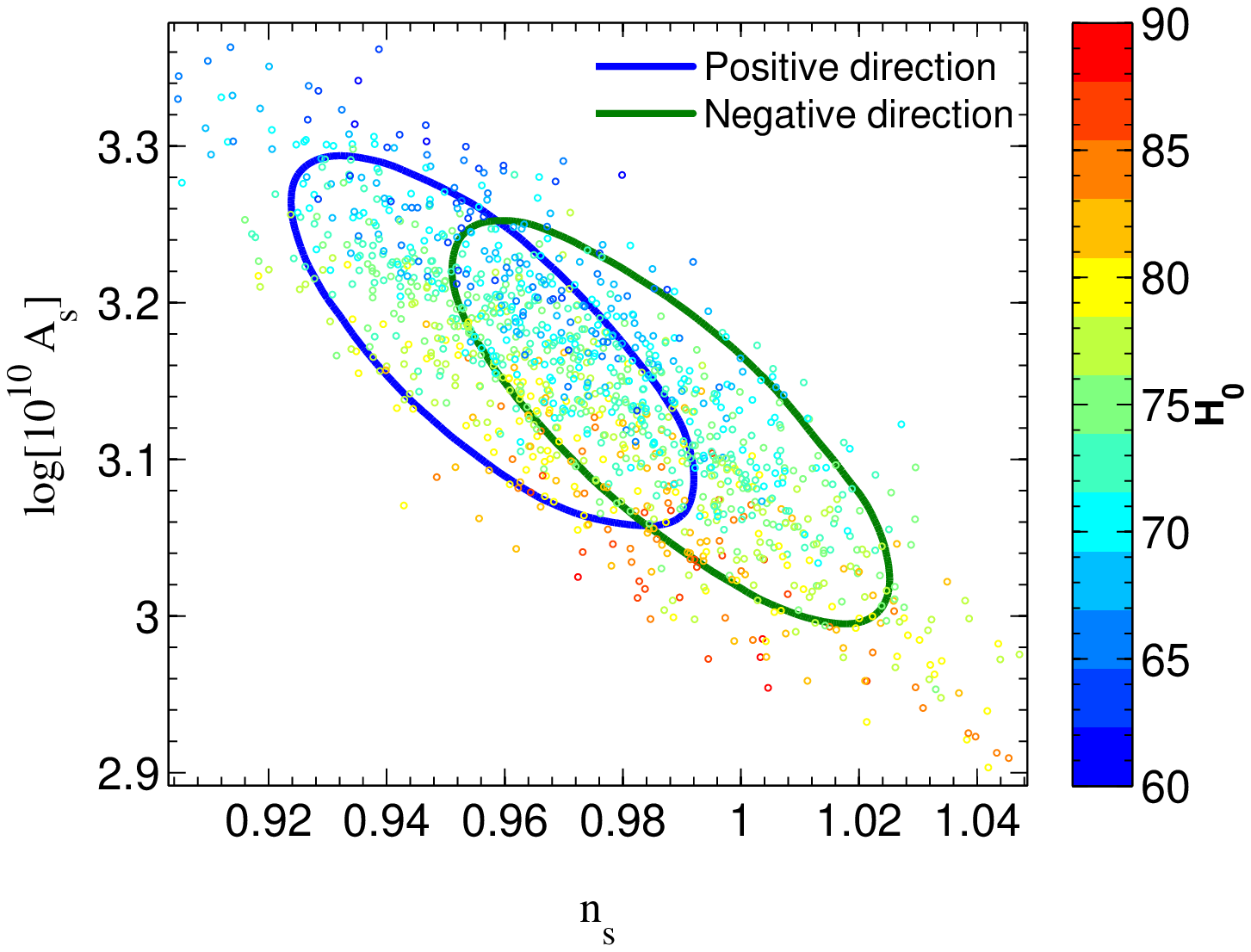}
\includegraphics[width=\linewidth,angle=0]{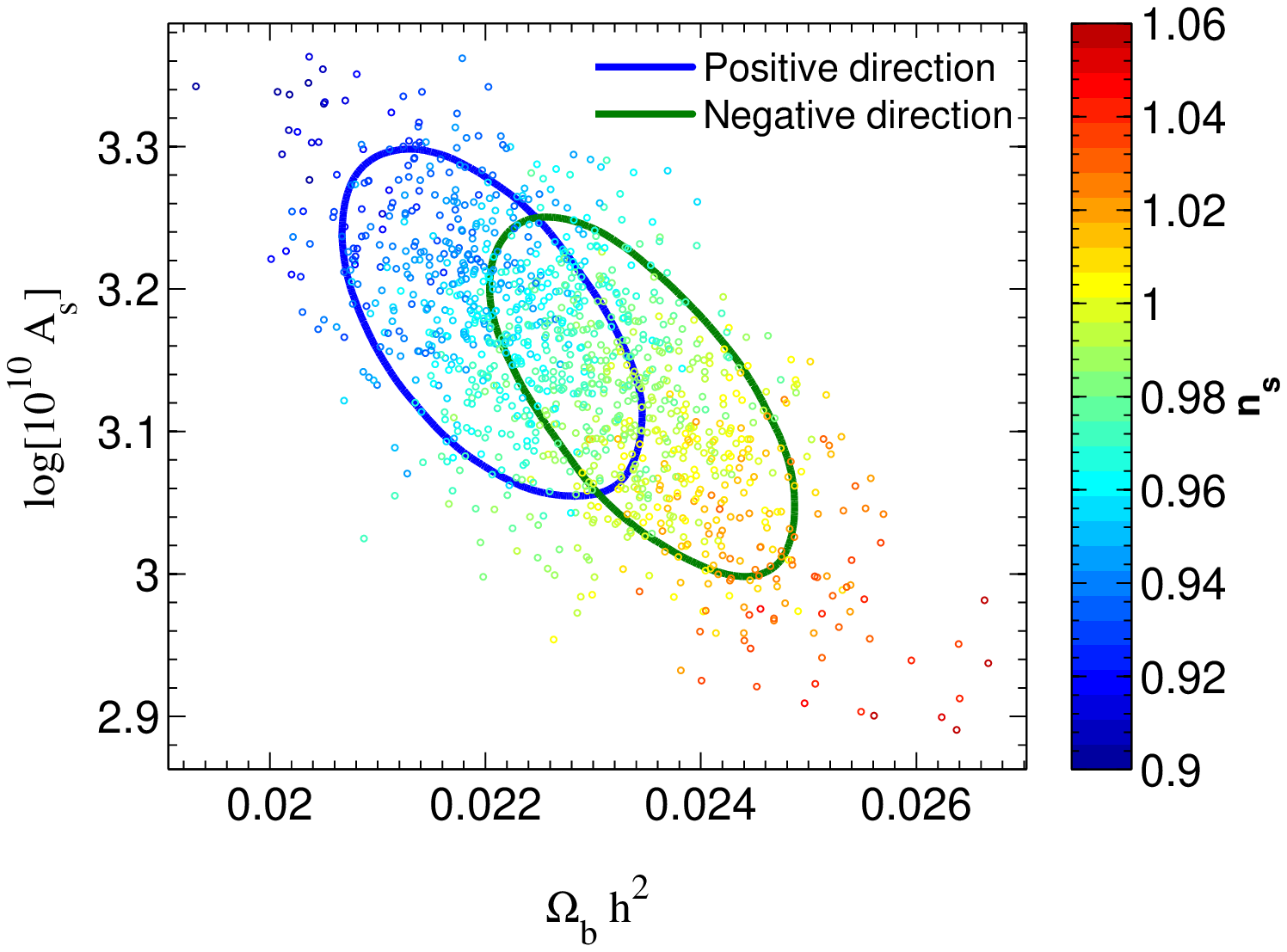}
\caption{\emph{Top}: Summary of the
  $n_{\textrm{s}}$--$\log(10^{10}A_{\textrm{s}})$ posterior in terms
  of $1\sigma$ contours for the 12 regions. \emph{Middle}: As above,
  but evaluated from antipodal hemispheres aligned with the
  preferred asymmetry direction. Colored dots indicate the values of
  $H_0$. \emph{Bottom}: As the middle plot, but for
  $\Omega_{\textrm{b}}h^2$--$\log(10^{10}A_{\textrm{s}})$.}
\label{fig:param2d}
\end{figure} 
  
For an isotropic map, the power spectrum should be uncorrelated
between multipoles. Although masking does introduce correlations
between adjacent multipoles, it is not expected that this will be
significant between the 100-multipole blocks, and therefore the dipole
directions should be random. This is confirmed by simulations.  The degree of
alignment between the dipole directions of different multipole blocks
is then used as a measure of the power spectrum asymmetry.

We also use the 100-multipole blocks to compute the power spectra for
the two opposite hemispheres defined by the direction of maximum
asymmetry, and for disks of diameter 90 degrees centered on the
same directions.

\paragraph{Cosmological parameter estimation:}
Our primary interest here is to evaluate the directional dependence of
the cosmological parameters in the temperature data.  We apply a
similar binning and power spectrum estimation method as described
above. For the high-$\ell$ likelihood, we compute the MASTER estimates
of the spectra from the full-resolution $\nside=512$ map.
For the low-$\ell$ likelihood, we replace the WMAP pixel likelihood by
a MASTER estimate computed for a single bin, $\ell \in [2,31]$. {
  Tests indicate that this modification does not introduce any
  significant deviation in the parameters on the full sky as compared to the
  official WMAP values. Indeed, the parameter estimate changes were
  insignificant when compared to each parameter's 1$\sigma$-value.}
The maximum multipole used in the parameter analysis is
$\ell_{\mathrm{max}}=1008$. For all spectra,
we subtract the best-fit unresolved point source amplitude
\citep{wmap92} before parameter estimation.
 
Our TT likelihood code 
uses the offset log-normal term that was introduced into the WMAP
likelihood in \citet{verde2003} - hence the total likelihood is a
linear combination of Gaussian and log-normal terms: \beq
-\log\mathcal{L}(C_b|\widehat{C}_b) \sim \frac{1}{3}\sum_{b,b'} \Delta
\mcb C^{-1}_{bb'} \Delta \mcbp^{T} + \frac{2}{3}\sum_{b,b'} \Delta z_b
\mathcal{Q}_{bb'} \Delta z_{b'}^{T} \eeq where $\widehat{C}_b$
and $C_b$ are the estimated and model power spectra respectively,
$\Delta \mcb = \mcb-\widehat{\mathcal{C}}_b$, $C^{-1}_{bb'}$ is the
covariance matrix, estimated using CMB plus noise MC simulations,
$z_b = \ln(C_b + N_b)$, (where $N_b$ is the noise spectrum)
and $\mathcal{Q}_{bb'} =
(\widehat{C}_b+N_b)C^{-1}_{bb'}({\widehat{C}_{b'}}+N_{b'})$ is the
local transformation of the covariance matrix to the log-normal
variables $z_b$. { The last term is added since a simple Gaussian
likelihood does not capture the full likelihood surface. A linear
combination of Gaussian+Log-normal terms has been tested and proven
to be minimally biased by the WMAP team \citep{verde2003}. The transformation 
to $z_b$ variables introduces no extra bias in the variance by construction,
and this implies the stated relationship between $C^{-1}$ and the curvature
matrix $\mathcal{Q}$. For further details see \citet{bond98}.}

To construct the covariance matrix, we use 10\,000 CMB plus noise Gaussian
simulations. The covariance matrix propagates the uncertainties
introduced by effects such as the noise, mask geometry, and associated
sample variance. We make no attempt to include beam uncertainties in our
pipeline. 


 The fractional areas of the 12 $\nside=1$ patches range from
$f_{\mathrm{sky}} = 0.019$ in the Galactic center to $f_{\mathrm{sky}}
= 0.085$ in regions at high latitude. With such small patches, one
might be concerned about the correctness of our likelihood
approximation. We have confirmed that the parameter estimates are
unbiased, performing parameter estimation on some of the small regions
in 500 simulated maps with known input parameters.

The final posterior distribution is comprised of the product of the
likelihood and a prior distribution that describes our previous
knowledge of the parameters. Since we do not consider polarization in
our analysis, we adopt a strong Gaussian prior on the reionization
redshift, $z_{\mathrm{rei}}=10.6\pm1.1$, which corresponds to the
WMAP9 best-fit value.  Due to the strong correlation of
$z_{\mathrm{re}}$ and the reionization optical depth $\tau$, we can
also obtain an additional constraint on this parameter as well.  { 
  We also used the \cosmomc default hard-coded priors on the Hubble
  constant and age of the Universe. The flat priors used in the other
  cosmological parameters is wide enough that the final estimates are
  dominated by the data.}

\section{Results}
\label{sect:results}

\paragraph{Power asymmetry: } 

In \fig\ref{fig:dipoles}, we show { the dipole directions of the
  first six 100-multipole bands as estimated from the corresponding $\nside=1$
  maps constructed from the 12 local power spectrum
  estimates}, together with the dipole for the full multipole range
$\ell=2-600$. These directions are consistent with those found from
the WMAP1 \citep{hansen2004} and WMAP5 \citep{hansen2009}
data sets, which are also indicated in the figure.

For a statistically isotropic CMB temperature distribution, the dipole
directions from uncorrelated power spectrum estimates
should be distributed randomly on the sky.  To quantify the
significance of the power asymmetry, we consider the dispersion angle,
which is defined as the mean angle, $\theta_\mathrm{mean}$, between
all possible combinations of 100-multipole dipole directions up to a
given $\ell_\mathrm{max}$.  The expected dispersion angle for Gaussian
simulations is 90 degrees, as confirmed by simulations.  We calculate
$\theta_\mathrm{mean}(\ell_\mathrm{max}=600)$ for the WMAP9 data and
compare it to the distribution obtained from 10\,000 CMB plus noise
simulations. { We found that only 7 of these exhibited a lower
  dispersion angle, implying a $3.4\sigma$ significance for the power
  asymmetry.}  This is lower than previously reported in
\cite{hansen2009}, where none of the 10\,000 simulations had a
similarly large mean angle.  However, there are several changes in
this analysis; (1) the new 9-year Galactic and point source masks
remove a larger fraction of the sky, thus increasing the scatter on
the dipole directions due to increased sample variance; (2) larger
disks are now used since the smaller disks from the 5-year analysis
result in several patches near the Galactic center with an extremely
small sky fraction when combined with the new Galactic mask; and (3)
12 independent regions are now used instead of 3072 overlapping ones
to speed-up the computations.  It is also interesting to note that we
have tried a number of permutations of the individual yearly sky maps
and found variations depending on which particular years were
excluded. However, the results always remain highly significant, and
the variations are found to be consistent with those determined from
simulations.

{ In order to illustrate the effect of the asymmetry on the power
  spectrum, we show in \fig\ref{fig:cldiff} the ratio $\Delta
  C_\ell/C_\ell$ of the power spectrum difference between the two
  antipodal hemispheres compared to their mean spectrum as computed
  along the maximum asymmetry direction. The olive-green band shows
  the $68\%$ confidence limit determined from simulations. Note that
  the corresponding mean ratio is larger than zero since the maximum
  asymmetry direction for each single simulation is used.  The
  amplitude of the mean ratio over the range $\ell=2-600$ for the data
  is exceeded in only $0.52\%$ of the simulations.

  Inspection of \fig\ref{fig:dipoles} suggests that some of the dipole
  directions are close to the south ecliptic pole. If the asymmetry
  had its origin in an instrumental systematic effect, or from some
  local foreground in the Solar System, then one might expect an
  alignment of the asymmetry with the ecliptic axis.  For this reason,
  we study this relation further. Note first that the distance from
  the direction of maximum asymmetry to the south ecliptic pole is 44
  degrees. We calculated the mean distance of the 6 dipole directions
  to the ecliptic pole as well as to the axis of maximum asymmetry and
  compared to simulations. While the mean angular distance to the
  direction of maximum asymmetry is smaller than that for the data in
  only $0.02\%$ of the simulations, the equivalent distance to the
  ecliptic pole is smaller in $3\%$ of the simulations. Furthermore,
  the significance of the $C_\ell$ ratio measured in opposing ecliptic
  hemispheres is $29\%$. We therefore conclude that the asymmetry is
  most likely not related to the ecliptic frame.

  Another possible mechanism for generating asymmetry is through the Doppler
  boosting of the CMB fluctuations due to our motion with respect to
  the CMB reference frame \citep[see][and references
  therein]{dopplerpaper}. This boosting causes a dipolar modulation of
  the amplitude of the fluctuations and a corresponding hemispherical asymmetry on all
  scales, and has been observed by \textit{Planck}.  It is therefore a
  strong candidate to produce an alignment of power dipoles as claimed
  here. However, using simulations we have determined that the
  magnitude of this effect at the WMAP frequencies is too small to
  have any impact on the hemispherical asymmetry described in this
  Letter. The power spectra in opposing hemispheres are changed by a
  maximum of $0.1\%$ and the mean dipole direction over the range
  $\ell=2-600$ is changed by only one degree. 

  Therefore, we consider if this asymmetry in power is reflected in fits to
  the standard \lcdm cosmological parameters.  }

\paragraph{Cosmological parameter estimation:}
\fig\ref{fig:lcdmpat} shows the directional dependence,
as specified by the 12 regions on the sky defined in
Figure~\ref{fig:regions} for the six main
\lcdm parameters.
The computed values and their standard deviations
are shown in black, whilst the corresponding results from our WMAP9
analysis on the full sky with the KQ85 mask applied is shown as a grey band.  
Inspecting the plots carefully, one finds that
the majority of parameter estimates fall within $\sim$$1\sigma$
of the WMAP9 full sky value. { One exception is for pixel 4 where
there is a $\sim3\sigma$ deviation for some parameters. This outlier
might be explained by residual foregrounds close
to the edge of the mask, or could simply be a large fluctuation.}

In \fig\ref{fig:dipoles}, we also show the
dipole directions of the $\nside=1$ parameter maps. Clearly $n_s$,
$A_s$ and $\Omega_b$ seem to show a directional dependence similar to
the power spectrum asymmetry and these seem to be the parameters
mostly affected by the asymmetry. {Note that the fitted dipole directions for
these parameters are only weakly affected by the outlier in pixel 4.}

In the left top panel of
\fig\ref{fig:param2d}, we demonstrate the $A_s - n_s$ correlation with $1\sigma$
contours for each region. All contours are consistent with each other
at better than 2$\sigma$, but some directional dependence is visible.

We also estimated parameters using hemispheres corresponding to the
preferred power asymmetry direction for $\ell=2-600$. Initially, we
restricted the analysis to $\lmax = 608$ to cover only that part of
the spectrum which is highly signal dominated and where the asymmetry
is prominent.  However, the absence of higher multipoles leads to
large uncertainties in the parameters of interest. We therefore
repeated the analysis for $\lmax=1008$. In this case, the error ellipses for $A_s$ vs
$\Omega_bh^2$ and $A_s$ vs $n_s$ computed on the positive
(power-enhanced) and negative (power-deficit) hemispheres show a
slight relative shift, as shown in the bottom panels of
\fig\ref{fig:param2d}. The best-fit parameters for each hemisphere lie
just at the border or the $1\sigma$ contours from the opposite
hemisphere. The two maximally asymmetric hemispheres do not,
therefore, indicate parameter values significantly different from the
WMAP9 full-sky results. It is interesting to note that the
power-deficit hemisphere prefers in general a higher $H_0$.  The
marginalised value obtained for the scalar spectral index in the two
hemispheres is $n_s = 0.959\pm 0.022$ and $n_s = 0.989 \pm 0.024$
respectively (an $\approx1.3\sigma$ difference). Note that in one
hemisphere, the spectral index is different from 1 at almost $2\sigma$
whereas in the other it is fully consistent with 1.

{In addition,  we compared the difference in parameter estimates
between the two opposite maximally asymmetric hemispheres of the
data to the corresponding \cosmomc estimates in 100 isotropic simulations.
In this way we were able to obtain a significance of asymmetry
for each single parameter. We found that the p-values for asymmetry in 
$\Omega_b h^2$, $\Omega_{DM} h^2$, $n_s$,
  $log[10^{10} A_s]$, $\theta$, $H_0$ are 32\%, 43\%, 53\% ,39\%,
  56\%, and 68\%, respectively, confirming that no asymmetry is
seen in the parameters. }

\section{Conclusions}

We measure a statistically significant power spectrum asymmetry in the
WMAP9 temperature sky maps with $3.4\sigma$ significance as measured
by the mean dispersion among the preferred directions derived from six
(nearly) independent multipole ranges between $\ell=2$ and 600, using
the conservative KQ85 mask adopted in the WMAP9 analysis. Only 7 out
of 10\,000 simulations show a similarly strong alignment.  The average
preferred direction points toward Galactic coordinates
$(l,b)=(226,-27)$, $44^\circ$ away from the south ecliptic pole
arguing against the possibility of a systematic or local astrophysical
cause for the asymmetry related to the ecliptic frame of reference.
Conversely, the cosmological parameters do not show a strong regional
dependence, although the parameters $A_s$, $n_s$ and $\Omega_b$ do
hint at a weak sensitivity to the hemispherical power asymmetry.

\begin{acknowledgments}
FKH acknowledges OYI grant from the Norwegian research council. HKE
acknowledges supported through the ERC Starting Grant
StG2010-257080. We acknowledge the use of resources from the Norwegian
national super computing facilities NOTUR. Maps and results have been
derived using the \healpix (http://healpix.jpl.nasa.gov) software
package developed by \cite{Healpix2005}. Results have been derived
using the \cosmomc code from \cite{Lewis2002}. We acknowledge the use
of the LAMBDA archive (Legacy Archive for Microwave Background Data
Analysis). Support for LAMBDA is provided by the NASA office for Space
Science.
\end{acknowledgments}


\end{document}